# A Phenomenological Thermodynamic Potential for CaTiO$_3$ Single Crystal

Yijia Gu[1], Karin Rabe[2], Eric Bousquet[3,4], Venkatraman Gopalan[1], and Long-Qing Chen[1]

[1]Department of Materials Science and Engineering, Pennsylvania State University, University Park, Pennsylvania 16802, USA

[2]Department of Physics and Astronomy, Rutgers University, Piscataway, New Jersey 08854-8019, USA

[3]Department of Materials, ETH Zürich, Wolfgang-Pauli-Strasse 27, CH-8093 Zürich, Switzerland

[4]Physique Théorique des Matériaux, Université de Liège, B-4000 Sart Tilman, Belgium

**Abstract**

The antiferrodistortive (AFD) structural transitions of calcium titanate (CaTiO$_3$) at ambient pressure have been extensively studied during the last few years. It is found none of the AFD polymorphs is polar or ferroelectric. However, it was recently shown theoretically and later experimentally confirmed that a ferroelectric transition in CaTiO$_3$ can be induced by tensile strains. The ferroelectric instability is believed to be strongly coupled to the AFD soft modes. In this article, we present a complete thermodynamic potential for describing the coupling between the AFD and ferroelectric phase transitions. We analyzed the dependence of transition temperatures on stress and strain condition. Based on this potential, a (001) CaTiO$_3$ thin film diagram was constructed. The results show good agreement with available experimental observations. The strong suppression of ferroelectric transition by the AFD transition is discussed.

## I. INTRODUCTION

The ideal perovskite structure, described as a simple cubic network of corner linked BO$_6$ octahedra with A atoms occupying 12-fold oxygen coordinated sites, is inherently unstable and can exhibit a variety of distortions. These include polar distortions, dominated by off-centering of B cation in its oxygen octahedron, and tilts and rotations of the oxygen octahedron network. The polar distortions lead to the presence of dipoles and to ferroelectric and antiferroelectric behavior in several well-known perovskite compounds such as BaTiO$_3$, PbTiO$_3$, PbZrO$_3$, and BiFeO$_3$.[1] Oxygen octahedron rotations produce a variety of nonpolar phases, the phase transitions of which are called antiferrodistortive (AFD) phase transitions. The same compound can show instabilities to both distortions in the cubic phase, in which case they usually compete. Strontium titanate (SrTiO$_3$) is a good example of such compounds. Although SrTiO$_3$ has a ferroelectric instability, it is paraelectric all the way down to 0 K. Its ferroelectric transition is suppressed by the proceeding AFD phase transition[2,3]. With a sufficiently large epitaxial strain, SrTiO$_3$ becomes a ferroelectric even at room temperature[4].

At ambient temperature and pressure, calcium titanate ($CaTiO_3$) has the orthorhombic distorted-perovskite structure with space group *Pbnm*, a structure common to many perovskite oxides. Disregarding the distortion of $TiO_6$ octahedra, the structure of $CaTiO_3$ can be illustrated as a combination of two kinds of $TiO_6$ octahedron tilts: two out-of-phase tilts along $x_1$ and $x_2$ directions, and one in-phase tilt along $x_3$ direction (Fig. 1). With the standard Glazer's notation[5], it can be expressed as $a^-a^-c^+$. These two kinds of tilts can also be used to characterize the AFD transitions in $CaTiO_3$. We will discuss it in more details later.

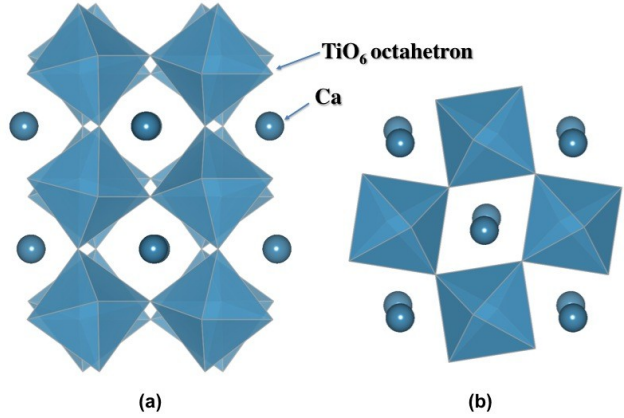

Figure 1. Crystal structure of $CaTiO_3$: (a) projection along [100] direction, the $TiO_6$ octahedra in consequtive two layers exhibit out-of-phase tilt, (the projection along [010] direction is similar); (b) projection along [001] direction, the $TiO_6$ octahedra show in-phase tilt.

The AFD transition sequence of $CaTiO_3$ is complicated. From high to low temperature, $CaTiO_3$ transforms from cubic ($Pm\bar{3}m$) to tetragonal (*I4/mcm*) at about 1600 K, and from tetragonal (*I4/mcm*) to orthorhombic at about 1500 K. [6-11] The later transition or transitions is quite controversial. Ali and Yashima[10, 11] proposed a direction transition from *I4/mcm* to *Pbnm* by the Rietveld analysis of high temperature X-ray and neutron diffraction data. Also by the analysis of high temperature neutron diffraction data, Kennedy[9] found there might be an intermediate phase with *Cmcm* structure between the transition from *I4/mcm* to *Pbnm*. And the transition temperature from *Cmcm* to *Pbnm* is around 1380 K, which agrees with both Guyot's drop-calorimetry measurements[7] and Gillet's Raman spectroscopy observation [12]. On the other hand, Carpenter theoretically investigated the structural transitions of $CaTiO_3$ using Landau theory, and he concluded that in order to get a stable *Pbnm* structure, there must be some intermediate structure between *I4/mcm* and *Pbnm*. However, he proposed an *I4/mcm*→ *Imma* → *Pbnm* transition sequence.

Despite of the complicity and discrepancy, none of the above mentioned structures is polar or ferroelectric at ambient pressure. However, $CaTiO_3$ has a ferroelectric soft mode as manifested by a high dielectric constant at low temperature [13] and later first principles calculations[14]. Experiments also show frequency independence of $CaTiO_3$ dielectric constants, which makes it a high-quality microwave material. Therefore, similar to $SrTiO_3$, $CaTiO_3$ is also an incipient ferroelectric[13], and the extrapolated ferroelectric transition temperature is about -111 K[13, 15]. It is natural to consider the ferroelectricity of $CaTiO_3$ as an analog to that of SrTiO$_3$, which is suppressed by AFD, but can be induced by applied strain.[2-4, 16] In addition, some other perovskites with *Pbnm* structures, including CaMnO$_3$,[17] SrZrO$_3$,[18] and etc[19], are possible to exhibit strain-induced ferroelectricity. Recently, by first principles calculations Eklund et al[20, 21] predicted that 1.5% epitaxial tensile strain can indeed lead to ferroelectric transition. Experimentally, Vlahos et al[22] found spontaneous polarization in the CaTiO$_3$/NdGaO$_3$ film system with a tensile constraint strain of 1.15%. Thus, ferroelectricity in CaTiO$_3$ can be induced by a sufficiently large tensile strain.

In addition to the strain-induced ferroelectric behavior of thin films, the twin walls of CaTiO$_3$ have been extensively investigated, including trapping of oxygen vacancies[23, 24], the activation energy for twin-wall motion[25], and the intrinsic elasticity of the twin walls[26]. By theoretical simulations, Goncalves-Ferreira et al[27] showed that the CaTiO$_3$ ferroelastic twin walls exhibit sizeable spontaneous polarization due to the vanishing of octahedra tilt and the decrease of the material density. Further experiments show that the twins of CaTiO$_3$ are ferroelectric themselves.[22] Since the formation of twins is usually to lower the total strain energy, the twins themselves are usually strained. Therefore, the discovered ferroelectricity of CaTiO$_3$ twin domains may also be due to strain effect.

In order to control and manipulate its properties with an applied external strain, it is necessary to understand the thermodynamics of CaTiO$_3$. Carpenter [28, 29] proposed a Landau expansion to describe the AFD transitions in (Ca, Sr)TiO$_3$. Although he made a systematic analysis of the stability of all the possible structures, the ferroelectric transition is not considered, and coefficients were not determined. In this paper, we construct a phenomenological thermodynamic potential for a CaTiO$_3$ single crystal, which incorporates both the AFD transitions and the ferroelectric transitions with different stress and strain conditions. This potential can therefore be employed to analyze all the important phase transitions and their dependence on stress and strain conditions.

## II. PHENOMENOLOGICAL DESCRIPTION

The phase transitions in CaTiO$_3$ can be described with a single Landau free energy expansion in terms of $\varepsilon_i$, $P_i$, and $q_i$. $\varepsilon_i$ (i = 1 – 6) are the strain components following Voigt's convention. $P_i$ (i=1, 2, 3) represent three components of the spontaneous polarization in the Cartesian coordinate system. $q_i$ (i=1, 2, 3) represent the linear oxygen displacement that corresponds to simultaneous out-of-phase tilt of TiO$_6$ octahedra. Similarly, $q_i$ (i=4, 5, 6) represent the oxygen displacement of simultaneous in-phase tilt of TiO$_6$ octahedra. The relationship between order parameter $q_i$ and octahedral tilt angles are explained in the Appendix. In terms of soft modes, $P_i$, $q_i$ (i=1, 2, 3), and $q_i$ (i=4, 5, 6) correspond to the $\Gamma_4^-$, $R_4^+$, $M_3^+$ modes, respectively. The total free energy has following form,

$$F = F_{Polar} + F_{OPT} + F_{IPT} + F_{Elastic} + F_{Coupling} \qquad (1)$$

The first three terms on the right-hand side of (1) describe contributions from spontaneous polarization, out-of-phase tilt, and in-phase tilt,

$$F_{Polar} = \alpha_1(T)(P_1^2 + P_2^2 + P_3^2) + \alpha_{11}(P_1^2 + P_2^2 + P_3^2)^2 + \alpha_{12}(P_1^4 + P_2^4 + P_3^4)$$
$$+ \alpha_{111}(P_1^2 + P_2^2 + P_3^2)^3 + \alpha_{112}(P_1^2 + P_2^2 + P_3^2)(P_1^4 + P_2^4 + P_3^4) + \alpha_{122}(P_1 P_2 P_3)^2 \quad (2)$$

$$F_{OPT} = \beta_1(T)(q_1^2 + q_2^2 + q_3^2) + \beta_{11}(q_1^2 + q_2^2 + q_3^2)^2 + \beta_{12}(q_1^4 + q_2^4 + q_3^4)$$
$$+ \beta_{111}(q_1^2 + q_2^2 + q_3^2)^3 + \beta_{112}(q_1^2 + q_2^2 + q_3^2)(q_1^4 + q_2^4 + q_3^4) + \beta_{122}(q_1 q_2 q_3)^2 \quad (3)$$

$$F_{IPT} = \gamma_1(T)(q_4^2 + q_5^2 + q_6^2) + \gamma_{11}(q_4^2 + q_5^2 + q_6^2)^2 + \gamma_{12}(q_4^4 + q_5^4 + q_6^4)$$
$$+ \gamma_{111}(q_4^2 + q_5^2 + q_6^2)^3 + \gamma_{112}(q_4^2 + q_5^2 + q_6^2)(q_4^4 + q_5^4 + q_6^4) + \gamma_{122}(q_4 q_5 q_6)^2 \quad (4)$$

where $\alpha$, $\beta$, and $\gamma$ are constants. Only the coefficients of the second order terms are assumed to be temperature dependent, i.e.

$$\alpha_1(T) = \alpha_{10}\Theta_{S1}[\coth(\frac{\Theta_{S1}}{T}) - \coth(\frac{\Theta_{S1}}{T_1})]$$
$$\beta_1(T) = \beta_{10}\Theta_{S2}[\coth(\frac{\Theta_{S2}}{T}) - \coth(\frac{\Theta_{S2}}{T_2})] \quad (5)$$
$$\gamma_1(T) = \gamma_{10}\Theta_{S3}[\coth(\frac{\Theta_{S3}}{T}) - \coth(\frac{\Theta_{S3}}{T_3})]$$

where $T_1$, $T_2$, and $T_3$ are Curie temperatures, $\Theta_{S1}$, $\Theta_{S2}$, and $\Theta_{S3}$ are saturation temperatures. The strain contribution to the total free energy can be written as

$$F_{Elastic} = \frac{1}{2}C_{11}(\varepsilon_1^2 + \varepsilon_2^2 + \varepsilon_3^2) + C_{12}(\varepsilon_1\varepsilon_2 + \varepsilon_3\varepsilon_2 + \varepsilon_1\varepsilon_3) + \frac{1}{2}C_{44}(\varepsilon_4^2 + \varepsilon_5^2 + \varepsilon_6^2) \quad (6)$$

where $C_{11}$, $C_{12}$, and $C_{44}$ are elastic stiffness constants; $\varepsilon_1$- $\varepsilon_6$ are strain components. The coupling energy among different order parameters and strains is written as

$$F_{Coupling} = -t_{11}(P_1^2 q_1^2 + P_2^2 q_2^2 + P_3^2 q_3^2) - t_{12}[P_1^2(q_2^2 + q_3^2) + P_2^2(q_1^2 + q_3^2) + P_3^2(q_1^2 + q_2^2)]$$
$$-t_{44}(P_1 P_2 q_1 q_2 + P_1 P_3 q_1 q_3 + P_2 P_3 q_2 q_3)$$
$$-\kappa_{11}(P_1^2 q_4^2 + P_2^2 q_5^2 + P_3^2 q_6^2) - \kappa_{12}[P_1^2(q_5^2 + q_6^2) + P_2^2(q_4^2 + q_6^2) + P_3^2(q_4^2 + q_5^2)]$$
$$-\kappa_{44}(P_1 P_2 q_4 q_5 + P_1 P_3 q_4 q_6 + P_2 P_3 q_5 q_6)$$
$$-\mu_{11}(q_1^2 q_4^2 + q_2^2 q_5^2 + q_3^2 q_6^2) - \mu_{12}[(q_2^2 + q_3^2)q_4^2 + (q_3^2 + q_1^2)q_5^2 + (q_1^2 + q_2^2)q_6^2]$$
$$-g_{11}(P_1^2 \varepsilon_1 + P_2^2 \varepsilon_2 + P_3^2 \varepsilon_3) - g_{12}[\varepsilon_1(P_2^2 + P_3^2) + \varepsilon_2(P_1^2 + P_3^2) + \varepsilon_3(P_1^2 + P_2^2)]$$
$$-g_{44}(P_1 P_2 \varepsilon_6 + P_1 P_3 \varepsilon_5 + P_2 P_3 \varepsilon_4)$$
$$-\lambda_{11}(\varepsilon_1 q_1^2 + \varepsilon_2 q_2^2 + \varepsilon_3 q_3^2) - \lambda_{12}[\varepsilon_1(q_2^2 + q_3^2) + \varepsilon_2(q_3^2 + q_1^2) + \varepsilon_3(q_1^2 + q_2^2)]$$
$$-\lambda_{44}(\varepsilon_4 q_2 q_3 + \varepsilon_5 q_3 q_1 + \varepsilon_6 q_1 q_2)$$
$$-\varsigma_{11}(\varepsilon_1 q_4^2 + \varepsilon_2 q_5^2 + \varepsilon_3 q_6^2) - \varsigma_{12}[\varepsilon_1(q_5^2 + q_6^2) + \varepsilon_2(q_6^2 + q_4^2) + \varepsilon_3(q_4^2 + q_5^2)]$$
$$-\varsigma_{44}(\varepsilon_4 q_5 q_6 + \varepsilon_5 q_6 q_4 + \varepsilon_6 q_4 q_5) \quad (7)$$

where $t_{ij}$, $\kappa_{ij}$, $g_{ij}$, $\mu_{ij}$, $\lambda_{ij}$, and $\varsigma_{ij}$ are coupling coefficients. Many of the parameters have already been calculated by first principles calculations[21]. The parameters converted from first principles calculations are listed in the Table 1.

Table 1. The parameters converted from first principles calculations*.[21] (energy density unit: J/m$^3$)

| | | | | | | | |
|---|---|---|---|---|---|---|---|
| $\alpha_1$ | -3.56×10$^8$ | $\beta_{111}$ | -2.89×10$^7$ | $C_{11}$ | 4.03×10$^{11}$ | $g_{11}$ | 1.02×10$^{10}$ |
| $\alpha_{11}$ | 3.70×10$^8$ | $\beta_{112}$ | -2.31×10$^8$ | $C_{12}$ | 1.07×10$^{11}$ | $g_{12}$ | -1.76×10$^9$ |
| $\alpha_{12}$ | 9.72×10$^7$ | $\beta_{122}$ | -4.92×10$^8$ | $C_{44}$ | 9.99×10$^{10}$ | $g_{44}$ | 7.70×10$^9$ |
| $\alpha_{111}$ | -1.18×10$^7$ | $\gamma_1$ | -1.85×10$^9$ | $t_{11}$ | -1.53×10$^9$ | $\lambda_{11}$ | -2.10×10$^9$ |
| $\alpha_{112}$ | -5.94×10$^7$ | $\gamma_{11}+\gamma_{12}$ | 1.48×10$^9$ | $t_{12}$ | -7.79×10$^8$ | $\lambda_{12}$ | -9.85×10$^9$ |
| $\alpha_{122}$ | -2.68×10$^8$ | $\gamma_{111}+\gamma_{112}$ | -2.31×10$^8$ | $t_{44}$ | 2.34×10$^9$ | $\lambda_{44}$ | -1.24×10$^9$ |
| $\beta_1$ | -2.05×10$^9$ | $\gamma_{122}$ | - | $\kappa_{11}$ | -1.43×10$^9$ | $\varsigma_{11}$ | 0 |
| $\beta_{11}$ | 1.20×10$^9$ | $\mu_{11}$ | -7.69×10$^9$ | $\kappa_{12}$ | -5.02×10$^8$ | $\varsigma_{12}$ | -9.65×10$^9$ |
| $\beta_{12}$ | 3.62×10$^8$ | $\mu_{12}$** | 3.29×10$^8$ | $\kappa_{44}$ | - | $\varsigma_{44}$ | - |

* $R_5^+$ mode is neglected;

** Normalized by eliminating $X_5^+$ mode.

## III. RESULTS AND DISCUSSION

### A. AFD transitions

For the AFD transition with only one in-phase TiO$_6$ octahedron tilt and two out-of-phase TiO$_6$ octahedron tilts, i.e. $P_1=P_2=P_3=q_3=q_4=q_5=0$, we have

$$F = \beta_{10}\Theta_{S2}[\coth(\frac{\Theta_{S2}}{T}) - \coth(\frac{\Theta_{S2}}{T_2})](q_1^2 + q_2^2) + \gamma_{10}\Theta_{S3}[\coth(\frac{\Theta_{S3}}{T}) - \coth(\frac{\Theta_{S3}}{T_3})]q_6^2$$
$$+ \beta_{11}^*(q_1^2 + q_2^2)^2 + \beta_{12}^*(q_1^4 + q_2^4) + (\gamma_{11}^* + \gamma_{12}^*)q_6^4 - \mu_{12}^*(q_1^2 + q_2^2)q_6^2$$
$$+ \beta_{111}(q_1^2 + q_2^2)^3 + \beta_{112}(q_1^2 + q_2^2)(q_1^4 + q_2^4) + (\gamma_{111} + \gamma_{112})q_6^6$$

(8)

where $\beta_{ij}^*$, $\mu_{ij}^*$ and $\gamma_{ij}^*$ are normalized coefficients with stress-free boundary condition (see Appendix for detail). The order parameters and free energies of different structures are summarized in Table 2.

Table 2. The order parameters and free energies of different structures of AFD transitions.

| Space Group | Order Parameters | Energy Expression |
|---|---|---|
| $Pm\bar{3}m$ | $q_i=0$, (i=1, 2, 6) | 0 |
| $I4/mcm$ | $q_1 \neq 0$ | $F_{I4/mcm} = \beta_1(T)q_1^2 + \beta_{11}^* q_1^4 + (\beta_{111} + \beta_{112})q_1^6$ |
| $Imma$ | $q_1 = q_2 \neq 0$ | $F_{Imma} = 2\beta_1(T)q_1^2 + (2\beta_{11}^* + \beta_{12}^*)q_1^4 + (8\beta_{111} + 2\beta_{112})q_1^6$ |
| $Cmcm$ | $q_1 \neq q_6 \neq 0$ | $F_{Cmcm} = \beta_1(T)q_1^2 + \beta_{11}^* q_1^4 + (\beta_{111} + \beta_{112})q_1^6 + \gamma_1(T)q_6^2 + \gamma_{11}^* q_6^4 + (\gamma_{111} + \gamma_{112})q_6^6 - \mu_{12}^* q_1^2 q_6^2$ |
| $Pbnm$ | $q_1 = q_2 \neq 0$, $q_6 \neq 0$ | $F_{Pbnm} = 2\beta_1(T)q_1^2 + (2\beta_{11}^* + \beta_{12}^*)q_1^4 + (8\beta_{111} + 2\beta_{112})q_1^6 + \gamma_1(T)q_6^2 + \gamma_{11}^* q_6^4 + (\gamma_{111} + \gamma_{112})q_6^6 - 2\mu_{12}^* q_1^2 q_6^2$ |

According to experimental results as discussed in the introduction, we can conclude that there are at least two AFD transitions, i.e. $Pm\bar{3}m$ to $I4/mcm$, and another transition to $Pbnm$. The latter can't be a direct transition from $I4/mcm$ to $Pbnm$, if the energy of $Imma$ or $Cmcm$ is higher than $Pbnm$. As compared in Table 2, appropriate selection of coefficients can generate different possibilities for the latter AFD transition sequence, such as $I4/mcm \rightarrow Imma \rightarrow Pbnm$, $I4/mcm \rightarrow Cmcm \rightarrow Pbnm$, and etc. Carpenter[29] analyzed the energy difference between these structures and proposed an $I4/mcm \rightarrow Imma \rightarrow Pbnm$ transition sequence. It should be noted that the $Imma$ structure was not observed experimentally. Here, we propose another scenario for the transformation sequence, $I4/mcm \rightarrow Cmcm \rightarrow Pbnm$, although the existence of $Cmcm$ structure is still controversial in this system [7, 9-11]. However, only this transition sequence can account for both the transition temperature of about 1380 K, which was determined by Guyot[7], and Gillet[12] respectively, and Kennedy's neutron diffraction results[9]. According to Guyot's heat capacity measurement [7], both $I4/mcm \rightarrow Cmcm$ and $Cmcm \rightarrow Pbnm$ transitions are of the first order. For the $Pm\bar{3}m \rightarrow I4/mcm$ transition at about 1600 K, there is no or very small latent heat, which may be buried by the broad calorimetric peak of the previous transition [7]. Therefore, this transition may be of the second order or weakly first-order. However, the tilt angles versus temperature diagram from the X-ray diffraction and neutron diffraction results[9,11] shows discontinuity near the transition temperature, a characteristic feature of a first-order transition.

In this paper, we adopted Guyot's [7] measured data of the transformation latent heat, and assumed that the $Pm\bar{3}m \rightarrow I4/mcm$ transition is also of first order with a small latent heat of 1.0 kJ/mol. The saturation temperatures were estimated from the (Ca,Sr)TiO$_3$ phase diagrams[30]. The calculated values of $\beta_{10}$ and $\gamma_{10}$ by first principles show good agreement with the measured latent heat. So we simply adopted them to make the whole set of parameters consistent. The other parameters were determined by fitting Kennedy's [9] and Yashima's [11] neutron diffraction and X-ray diffraction data. A comparison between the fitted parameters and those from first principles is shown in Table 3.

Table 3. Parameters from fitting and their counterparts from first-principles calculations.

| Parameters | $\Theta_{S2}$(K) | $\Theta_{S3}$(K) | $\beta_{10}$ | $\gamma_{10}$ | $\beta_{11}^*$ | $\beta_{111}$ | $\beta_{12}^*$ | $\beta_{112}$ | $\gamma_{11}^* + \gamma_{12}^*$ | $\gamma_{111} + \gamma_{112}$ |
|---|---|---|---|---|---|---|---|---|---|---|
| From fitting | 274 | 345 | - | - | $-1.41 \times 10^8$ | $1.45 \times 10^9$ | $-3.59 \times 10^8$ | $1.15 \times 10^9$ | $-3.38 \times 10^9$ | $1.15 \times 10^{10}$ |
| From first principles | - | - | $1.54 \times 10^6$ | $1.68 \times 10^6$ | $1.10 \times 10^9$ | $-2.89 \times 10^7$ | $2.64 \times 10^8$ | $-2.31 \times 10^8$ | $1.27 \times 10^9$ | $-2.31 \times 10^8$ |

As shown in Table 3, the fitted parameters deviate from those calculated by first principles. Both signs and magnitudes are different in almost every case. However, this can be expected because the first-principles is for 0 K and our fits are from the whole temperature range. The validity of the first-principles calculations can be tested by comparing the total free energy at 0 K from both sets of parameters. Actually, the difference is about 6.5% of the total free energy. Considering the possible errors and approximations made during the two calculations, this difference is small. In addition, the discrepancy is only confined to the parameters of the fourth and sixth order terms. The nice agreement between our fitted plot and the measured values (Fig. 2) indicates the accuracy of the parameters of the second order terms and coupling terms from first-principles. As shown in Fig. 2, the fitted plot not only reproduces three first-order transitions, but also shows the saturation of tilt angles at very low temperature. We also compared the free energy of these structures to study the phase stabilities, as plotted in Fig. 3. Although the differences between *I4/mcm* and *Imma* and between *Cmcm* and *Pbnm* are very small, the relative phase stability of different structures is just as we expected. And the small energy difference between *Cmcm* and *Pbnm* indicates the difficulty to get stable *Cmcm* phase during *in situ* X-ray diffraction and neutron diffraction experiments.

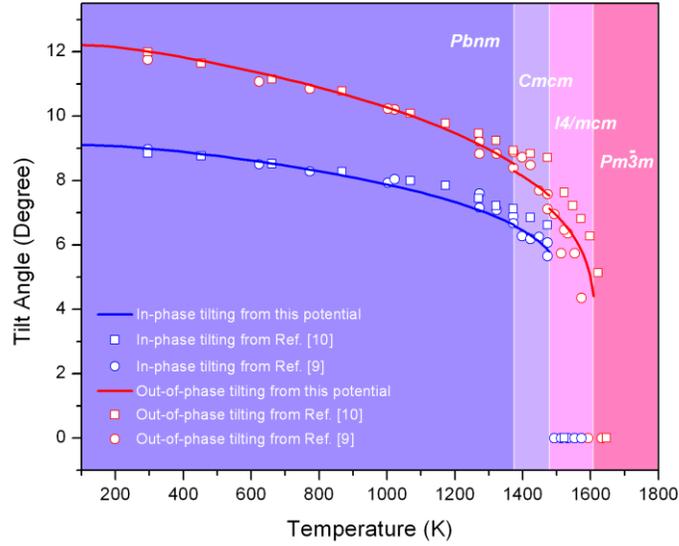

Figure 2. Tilt angle as a function of temperature. There discontinuities in the plot clearly show that there are three first order transformations.

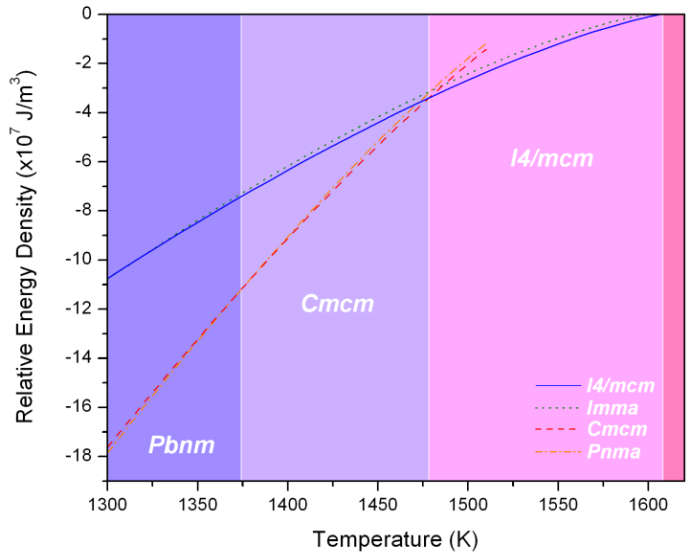

Figure 3. Relative free energy density of different structures: *I4/mcm*, *Imma*, *Cmcm*, and *Pbnm*. Note that *Pm$\bar{3}$m* is set to be the reference state with free energy equal to zero. So the relative free energies of other structures are basicly the energy difference from *Pm$\bar{3}$m* structure.

## B. Ferroelectric transition

With the refined parameters, we can further investigate the AFD effect on ferroelectric transition in CaTiO$_3$ single crystals. Firstly, we can extract the Curie temperature $T_1$ from the extrapolated value (-111 K)[13] by eliminating the coupling effect from TiO$_6$ octahedron tilts. From our model, it is easy to calculate the $T_1$'s for all the combinations of polarization in the three directions. And the calculated highest $T_1$ corresponds to the extrapolated ferroelectric effective temperature (-111 K).

By minimizing the free energy of AFD part, we can calculate the in-phase tilt angle and out-of-phase angle as $\varphi_3$=9.10° and $\theta_1$=$\theta_2$=8.64° respectively. Then, using the tilt angles and the saturation temperature $\Theta_{S3}$=55 K[13], the $T_1$'s of different polarization combinations are calculated. As listed in Table 4, the highest Curie temperature is 252.1 K for the case of $P_1$=$P_2$≠$P_3$. This structure is therefore the most stable one, and this temperature is the Curie temperature $T_1$. Correspondingly, the parameter $\alpha_{10}$ is calculated as 1.77×10$^6$.

Table 4. The calculated Curie temperatures for different polarization symmetry

| Polarization | $P_1$≠$P_2$≠$P_3$ | $P_1$=$P_2$≠$P_3$ | $P_1$=-$P_2$≠$P_3$ | $P_1$=$P_2$=$P_3$ |
|---|---|---|---|---|
| Curie temperature (K) | 139.4 | 252.1 | 132.0 | 187.4 |

Because the tilt angles do not change much at low temperature, we can simply freeze them and calculate the dielectric constant as a function of temperature. Thus, we get the coefficients of $P_1^2$ and $P_3^2$

$$\alpha_1^R = 2\alpha_{10}\Theta_{S1}[\coth(\frac{\Theta_{S1}}{T}) - \coth(\frac{\Theta_{S1}}{T_1})] - (2t_{11}^* + 2t_{12}^* + t_{44}^*)q_1^2 - 2\kappa_{12}^*q_6^2$$

$$\alpha_3^R = \alpha_{10}\Theta_{S1}[\coth(\frac{\Theta_{S1}}{T}) - \coth(\frac{\Theta_{S1}}{T_1})] - (t_{11}^* + t_{12}^*)q_1^2 - 2\kappa_{11}^*q_6^2$$

(9)

Experiment shows that the intensity of the optical second harmonic generation (SHG) of CaTiO$_3$ thin film changes continuously as a function of temperature[22], which indicates the ferroelectric transition of CaTiO$_3$ may be of second order. However, the defects in the thin films including strain inhomogeneity, domain structures, and so on, may make a first-order transformation look like a second order one. Further studies are needed to understand the nature of ferroelectric transition in CaTiO$_3$. In this paper, we assume the ferroelectric transformation of CaTiO$_3$ is second order. According to Devonshire's theory[1], the dielectric constant of a second order transformation can be written as

$$\varepsilon_{ij} = \frac{1}{\varepsilon_0 \alpha_{ij}} \qquad (i,j=1,2,3)$$

(10)

where $\varepsilon_0$ is the vacuum permittivity, and $\alpha_{ij}$ is the coefficient of $P_iP_j$ (i,j=1,2,3). Since $P_1$=$P_2$, it's easy to get $\varepsilon_{11} = \varepsilon_{22}$. The calculated dielectric constants are shown in Fig. 4. The total dielectric constant

($\sqrt{2\varepsilon_{11}^2 + \varepsilon_{33}^2}$) is 300 at 0 K, and 144 at room temperature. They are quite close to the measured values, 331 and 168[13], which indicate good accuracy for both the $\alpha_1$ value from first principles calculations and the Curie temperature $T_1$ from this calculation.

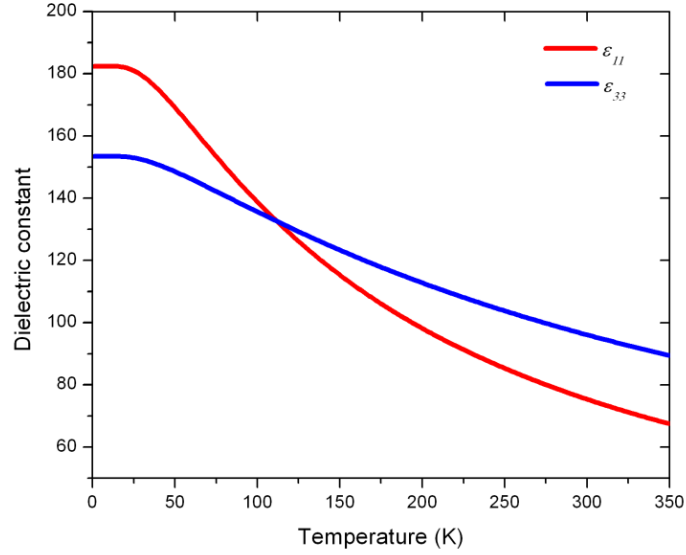

Figure 4. The dielectric constant as a function of temperature. The saturation of dielecric constants occurs at very low temperature.

With all the temperature-dependent coefficients, we can investigate the phase stability under different boundary conditions. Here we will calculate the temperature-constraint strain phase diagram of (001) CaTiO$_3$ thin film as an example.

For the stable structures of strained (001) CaTiO$_3$ thin films, Eklund[20, 21] reported two possible ferroelectric structures on the tensile strain side, *Pmc2$_1$* and *Pmn2$_1$*, among which *Pmn2$_1$* structure has slightly lower free energy. Also from first principles calculations, Bousquet[31] showed that *Pmc2$_1$* is stable. On the compressive side, *Pna2$_1$* is the stable structure.[21] In the following calculations, we will only consider these three structures.

Firstly, we renormalized the free energy expression with thin film boundary condition (see Appendix for detail). By minimizing the total free energy with respect to $q_1$ and $q_6$ respectively, we get

$$\beta_1'(T) + (2\beta_{11}' + \beta_{12}')q_1^2 + 6(2\beta_{111} + \beta_{112})q_1^4 - \mu_{13}'q_6^2 = 0 \qquad (11a)$$

$$\gamma_3'(T) + 2\gamma_{33}'q_6^2 + 3(\gamma_{111} + \gamma_{112})q_6^4 - 2\mu_{13}'q_1^2 = 0 \qquad (11b)$$

where $\beta_{ij}'$, $\mu_{ij}'$ and $\gamma_{ij}'$ are normalized coefficients. Combining (11a) and (11b) with equation from the coefficient of $P_1^2$,

$$2\alpha_1'(T) - (2t_{11}' + 2t_{12}' + t_{44})q_1^2 - 2\kappa_{13}'q_6^2 = 0 \tag{12}$$

we can get the phase boundary between *Pbnm* and *Pmc2₁* structures. It should be mentioned here, from our potential, the stable structure on the tensile side is *Pmc2₁*, not *Pmn2₁*. Similarly, for the phase boundary of *Pbnm*→ *Pna2₁* transition, we need to solve (11a), (11b), and the equation from the coefficient of $P_3^2$,

$$2\alpha_3'(T) - 2t_{31}'q_1^2 - 2\kappa_{33}'q_6^2 = 0 \tag{13}$$

The calculated phase diagram is asymmetric as shown in Fig. 5(a). The minimum tensile strain to induce the ferroelectric transition is about 1.5%, which agrees well with the prediction from the first principles calculations. On the compressive side of the diagram, about 13% compressive strain is needed to induce *Pbnm*→ *Pna2₁* transition. This value is so huge that it exceeds the limit of substrate constraint strain. In other words, it is impossible to have *Pna2₁* structure in (001) CaTiO₃ thin films. The temperature-constraint strain phase diagram of (001) CaTiO₃ thin film without AFD (Fig. 5(b)) was calculated by setting $q_i=0$ (i=1-6) and solving

$$\begin{vmatrix} \dfrac{\partial^2 F}{\partial P_1^2} & \dfrac{\partial^2 F}{\partial P_1 \partial P_2} & \dfrac{\partial^2 F}{\partial P_1 \partial P_3} \\ \dfrac{\partial^2 F}{\partial P_2 \partial P_1} & \dfrac{\partial^2 F}{\partial P_2^2} & \dfrac{\partial^2 F}{\partial P_2 \partial P_3} \\ \dfrac{\partial^2 F}{\partial P_3 \partial P_1} & \dfrac{\partial^2 F}{\partial P_3 \partial P_2} & \dfrac{\partial^2 F}{\partial P_3^2} \end{vmatrix}_{P_1=P_2=P_3=0} = 0 \tag{14}$$

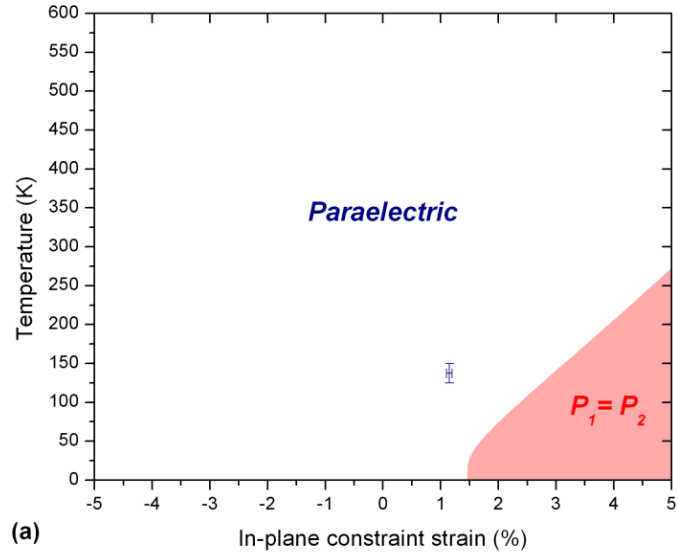

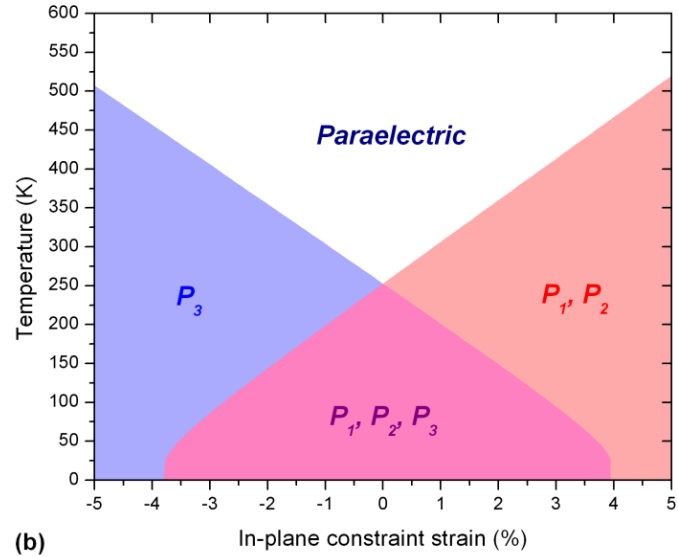

Figure 5. The temperature-constraint strain phase diagram of (001) $CaTiO_3$, (a) with AFD, and (b) without AFD. The transition point shown in (a) is measured by SHG experiment[22].

Comparing Fig. 5(a) and (b), we can easily find the asymmetry of the temperature-constraint strain phase diagram comes from the effect of AFD. Also the ferroelectric transition temperature of $CaTiO_3$ is greatly suppressed by AFD. A similar but weaker effect was also found in $SrTiO_3$[32]. The substantial effect of AFD on ferroelectric in $SrTiO_3$ is attributed to the competitive anharmonic couplings between AFD mode and ferroelectric mode, and their mutual coupling to the elasticity[2, 16]. In our phenomenological model of

$CaTiO_3$, the stability of different structures is strongly dependent on the coupling coefficients among $P_i$, $q_i$ and $\varepsilon_i$, which can be easily seen from equations (11a), (11b), (12) and (13). This indicates that the competition mechanism between AFD and ferroelectric is essentially the same as that of $SrTiO_3$.

By minimizing the total free energy, we also calculated the polarization of (001) $CaTiO_3$ thin film as a function of in-plane constraint tensile strain at different temperatures. As shown in Fig. 6, the ferroelectric transition temperature increases with in-plane tensile strain. At 0 K, the minimum tensile strain needed to induce the ferroelectric transition is about 1.5%. At 200 K the critical tensile strain increases to about 4% indicating the difficulty to obtain strain-induced ferroelectricity at elevated temperature. The calculated polarization of 4% tensile strain at 0 K is 0.61 $C/m^2$, which is more than twice that of $BaTiO_3$[1]. The polarization also exhibits saturation near the transition point, and becomes linear dependent on tensile strain in large strain region. As compared in the figure, our result of 0 K is a little larger than the first-principles calculations. The discrepancy may rise from different selection of stable structures. In the first-principles calculation[21], the stable structure used is *Pmc2₁* whereas we computed the polarization of *Pmn2₁*.

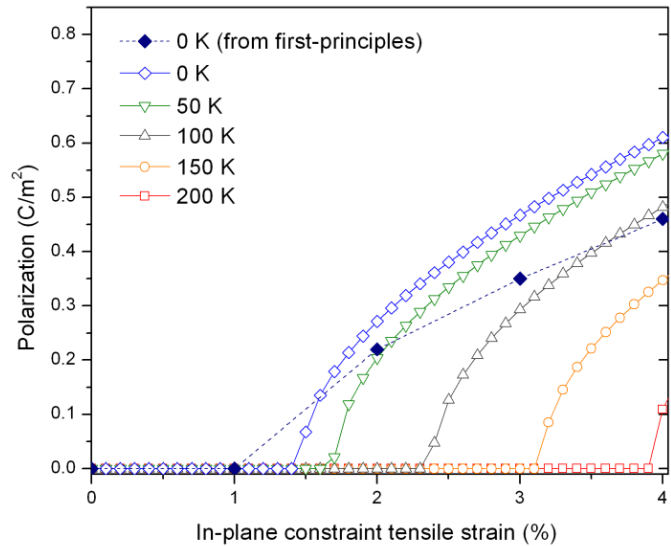

Figure 6. The poleriztation of (001) $CaTiO_3$ thin film as a function of in-plain constaint strain of different temperatures. The void markers represent our calculations for different temperatures. And the solid diamonds denote the data from first-principles calculations[21] (solid diamond) for 0 K.

## IV. CONCLUTIONS

A phenomenological thermodynamic potential is developed for $CaTiO_3$ single crystals. The coefficients of the potential are determined from first principles calculations and neutron diffraction and X-ray diffraction data. This potential effectively coupled the AFD transitions and strain-induced ferroelectric transitions. Several experimental observations, including transition temperatures, transition latent heat, dielectric constant, and tilt angles of TiO$_6$ octahedron, are successfully reproduced. Then the temperature-constraint strain single-domain phase diagram of (001) CaTiO$_3$ is constructed. The dependence of Curie temperature on constraint strain is quite asymmetric, i.e. only tensile strain can induce ferroelectric transition. Comparing the phase diagrams with and without AFD, we conclude that the asymmetry is not inherited from the ferroelectric transition itself but attributed from the AFD suppression.

**ACKNOWLEDGEMENT**

Y. Gu would like to thank Z. G. Mei for useful discussions. This work is supported by the NSF MRSEC under Grant No. DMR-0820404 and DMR-0908718.

**APPENDIX**

**1. Relationship between order parameter *q* and octahedral tilt angles**

For an infinitesimal angle, there is no octahedron distortion during tilting. The amplitude of $q_1$=1 means each atom move 1 Å along x$_1$ direction. Then in a simplified diagram of TiO$_6$ octahedron tilt, we have

$$\tan \theta_i = \frac{2 \times (q_i \times 0.5)}{a_0} = \frac{q_i}{a_0} \qquad (i = 1,2,3) \tag{A1}$$

where $\theta_i$ is the in-phase tilt angle, $a_0$ is the lattice parameter of the 5 atom cell. Similarly, we have the relationship for out-of-phase tilt

$$\tan \varphi_1 = \frac{q_i}{a_0} \qquad (i = 4,5,6) \tag{A2}$$

where $\varphi_i$ is the in-phase tilt angle

**2. Normalizing the total free energy with stress-free boundary condition**

With the stress-free boundary condition, we have

$$\frac{\partial F}{\partial \varepsilon_{ij}} = \sigma_{ij} = 0 \tag{A3}$$

Then we can rewrite the expression for the total free energy as

$$F = \alpha_{10}\Theta_{S1}[\coth(\frac{\Theta_{S1}}{T}) - \coth(\frac{\Theta_{S1}}{T_1})](P_1^2 + P_2^2 + P_3^2) + \alpha_{11}^*(P_1^2 + P_2^2 + P_3^2)^2 + \alpha_{12}^*(P_1^4 + P_2^4 + P_3^4)$$

$$+ \alpha_{111}(P_1^2 + P_2^2 + P_3^2)^3 + \alpha_{112}(P_1^2 + P_2^2 + P_3^2)(P_1^4 + P_2^4 + P_3^4) + \alpha_{122}(P_1 P_2 P_3)^2$$

$$+ \beta_{10}\Theta_{S2}[\coth(\frac{\Theta_{S2}}{T}) - \coth(\frac{\Theta_{S2}}{T_2})](q_1^2 + q_2^2 + q_3^2) + \beta_{11}^*(q_1^2 + q_2^2 + q_3^2)^2 + \beta_{12}^*(q_1^4 + q_2^4 + q_3^4)$$

$$+ \beta_{111}(q_1^2 + q_2^2 + q_3^2)^3 + \beta_{112}(q_1^2 + q_2^2 + q_3^2)(q_1^4 + q_2^4 + q_3^4) + \beta_{122}(q_1 q_2 q_3)^2$$

$$+ \gamma_{10}\Theta_{S3}[\coth(\frac{\Theta_{S3}}{T}) - \coth(\frac{\Theta_{S3}}{T_3})](q_4^2 + q_5^2 + q_6^2) + \gamma_{11}^*(q_4^2 + q_5^2 + q_6^2)^2 + \gamma_{12}^*(q_4^4 + q_5^4 + q_6^4)$$

$$+ \gamma_{111}(q_4^2 + q_5^2 + q_6^2)^3 + \gamma_{112}(q_4^2 + q_5^2 + q_6^2)(q_4^4 + q_5^4 + q_6^4) + \gamma_{122}(q_4 q_5 q_6)^2$$

$$- \mu_{11}^*(q_1^2 q_4^2 + q_2^2 q_5^2 + q_3^2 q_6^2) - \mu_{12}^*[(q_2^2 + q_3^2)q_4^2 + (q_3^2 + q_1^2)q_5^2 + (q_1^2 + q_2^2)q_6^2]$$

$$- t_{11}^*(P_1^2 q_1^2 + P_2^2 q_2^2 + P_3^2 q_3^2) - t_{12}^*[P_1^2(q_2^2 + q_3^2) + P_2^2(q_1^2 + q_3^2) + P_3^2(q_1^2 + q_2^2)]$$

$$- t_{44}^*(P_1 P_2 q_1 q_2 + P_1 P_3 q_1 q_3 + P_2 P_3 q_2 q_3) - \kappa_{11}^*(P_1^2 q_4^2 + P_2^2 q_5^2 + P_3^2 q_6^2)$$

$$- \kappa_{12}^*[P_1^2(q_5^2 + q_6^2) + P_2^2(q_4^2 + q_6^2) + P_3^2(q_4^2 + q_5^2)] - \kappa_{44}^*(P_1 P_2 q_4 q_5 + P_1 P_3 q_4 q_6 + P_2 P_3 q_5 q_6)$$

$$- \frac{\varsigma_{44}\lambda_{44}}{C_{44}}(q_2 q_3 q_5 q_6 + q_1 q_2 q_4 q_5 + q_1 q_3 q_4 q_6)$$

(A4)

where the * sign designates the renormalized coefficients, i.e.

$$\alpha_{11}^* = \alpha_{11} - \frac{C_{11}(g_{12}^2 + 2g_{11}g_{12}) - C_{12}(g_{11}^2 + 2g_{12}^2)}{2(C_{11} - C_{12})(C_{11} + 2C_{12})} - \frac{g_{44}^2}{4C_{44}},$$

$$\alpha_{12}^* = \alpha_{12} - \frac{(g_{11} - g_{12})^2}{2(C_{11} - C_{12})} + \frac{g_{44}^2}{4C_{44}},$$

$$\beta_{11}^* = \beta_{11} - \frac{C_{11}(\lambda_{12}^2 + 2\lambda_{11}\lambda_{12}) - C_{12}(\lambda_{11}^2 + 2\lambda_{12}^2)}{2(C_{11} - C_{12})(C_{11} + 2C_{12})} - \frac{\lambda_{44}^2}{4C_{44}},$$

$$\beta_{12}^* = \beta_{12} - \frac{(\lambda_{11} - \lambda_{12})^2}{2(C_{11} - C_{12})} + \frac{\lambda_{44}^2}{4C_{44}},$$

$$\gamma_{11}^* = \gamma_{11} - \frac{C_{11}(\varsigma_{12}^2 + 2\varsigma_{11}\varsigma_{12}) - C_{12}(\varsigma_{11}^2 + 2\varsigma_{12}^2)}{2(C_{11} - C_{12})(C_{11} + 2C_{12})} - \frac{\varsigma_{44}^2}{4C_{44}},$$

$$\gamma_{12}^* = \gamma_{12} - \frac{(\varsigma_{11} - \varsigma_{12})^2}{2(C_{11} - C_{12})} + \frac{\varsigma_{44}^2}{4C_{44}},$$

$$\mu_{11}^* = \mu_{11} + \frac{C_{11}(\lambda_{11}\varsigma_{11} + 2\lambda_{12}\varsigma_{12}) + C_{12}(\lambda_{11}\varsigma_{11} - 2\lambda_{11}\varsigma_{12} - 2\lambda_{12}\varsigma_{11})}{(C_{11} - C_{12})(C_{11} + 2C_{12})},$$

$$\mu_{12}^* = \mu_{12} - \frac{C_{12}(\lambda_{11}\varsigma_{11} + 2\lambda_{12}\varsigma_{12}) - C_{11}(\lambda_{12}\varsigma_{11} + \lambda_{11}\varsigma_{12} + \lambda_{12}\varsigma_{12})}{(C_{11} - C_{12})(C_{11} + 2C_{12})}.$$

$$t_{11}^* = t_{11} + \frac{C_{11}(\lambda_{11}g_{11} + 2\lambda_{12}g_{12}) + C_{12}(\lambda_{11}g_{11} - 2\lambda_{11}g_{12} - 2\lambda_{12}g_{11})}{(C_{11} - C_{12})(C_{11} + 2C_{12})},$$

$$t_{12}^* = t_{12} - \frac{C_{12}(\lambda_{11}g_{11} + 2\lambda_{12}g_{12}) - C_{11}(\lambda_{12}g_{11} + \lambda_{11}g_{12} + \lambda_{12}g_{12})}{(C_{11} - C_{12})(C_{11} + 2C_{12})},$$

$$t_{44}^* = t_{44} + \frac{\lambda_{44}g_{44}}{C_{44}},$$

$$\kappa_{11}^* = \kappa_{11} + \frac{C_{11}(\varsigma_{11}g_{11} + 2\varsigma_{12}g_{12}) + C_{12}(\varsigma_{11}g_{11} - 2\varsigma_{11}g_{12} - 2\varsigma_{12}g_{11})}{(C_{11} - C_{12})(C_{11} + 2C_{12})},$$

$$\kappa_{12}^* = \kappa_{12} - \frac{C_{12}(\varsigma_{11}g_{11} + 2\varsigma_{12}g_{12}) - C_{11}(\varsigma_{12}g_{11} + \varsigma_{11}g_{12} + \varsigma_{12}g_{12})}{(C_{11} - C_{12})(C_{11} + 2C_{12})},$$

$$\kappa_{44}^* = \kappa_{44} + \frac{\varsigma_{44}g_{44}}{C_{44}}.$$

(A5)

### 3. Normalizing the total free energy with thin film boundary condition

The thin film boundary condition is a mixed set of strain and stress boundary conditions. For (001) CaTiO$_3$ thin film, there is a biaxial strain in the $x_1$-$x_2$ plane, and all the stress components associated with $x_3$ direction are equal to zero, i.e.

$$\varepsilon_{11} = \varepsilon_{22} = \varepsilon_S, \quad \varepsilon_{12} = \varepsilon_{21} = 0,$$
$$\text{and} \quad \sigma_{13} = \sigma_{23} = \sigma_{31} = \sigma_{32} = \sigma_{33} = 0$$

(A6)

where $\varepsilon_S$ is the constraint strain. To satisfy the above stress-free condition it requires that

$$\frac{\partial F}{\partial \varepsilon_{ij}} = \sigma_{ij} = 0 \quad (ij = 13, 23, 31, 32, 33)$$

(A7)

So we have

$$\begin{aligned}
F = &\ \alpha'_1(T)(P_1^2 + P_2^2) + \alpha'_3(T)P_3^2 + \alpha'_{11}(P_1^4 + P_2^4) + \alpha'_{33}P_3^4 + \alpha'_{12}P_1^2P_2^2 + \alpha'_{13}(P_1^2 + P_2^2)P_3^2 \\
&+ \alpha_{111}(P_1^2 + P_2^2 + P_3^2)^3 + \alpha_{112}(P_1^2 + P_2^2 + P_3^2)(P_1^4 + P_2^4 + P_3^4) + \alpha_{122}(P_1 P_2 P_3)^2 \\
&+ \beta'_1(T)(q_1^2 + q_2^2) + \beta'_3(T)q_3^2 + \beta'_{11}(q_1^4 + q_2^4) + \beta'_{33}q_3^4 + \beta'_{12}q_1^2q_2^2 + \beta'_{13}(q_1^2 + q_2^2)q_3^2 \\
&+ \beta_{111}(q_1^2 + q_2^2 + q_3^2)^3 + \beta_{112}(q_1^2 + q_2^2 + q_3^2)(q_1^4 + q_2^4 + q_3^4) + \beta_{122}(q_1 q_2 q_3)^2 \\
&+ \gamma'_1(T)(q_4^2 + q_5^2) + \gamma'_3(T)q_6^2 + \gamma'_{11}(q_4^4 + q_5^4) + \gamma'_{33}q_6^4 + \gamma'_{12}q_4^2q_5^2 + \gamma'_{13}(q_4^2 + q_5^2)q_6^2 \\
&+ \gamma_{111}(q_4^2 + q_5^2 + q_6^2)^3 + \gamma_{112}(q_4^2 + q_5^2 + q_6^2)(q_4^4 + q_5^4 + q_6^4) + \gamma_{122}(q_4 q_5 q_6)^2 \\
&- \mu'_{11}(q_1^2q_4^2 + q_2^2q_5^2) - \mu'_{33}q_3^2q_6^2 - \mu'_{12}(q_2^2q_4^2 + q_1^2q_5^2) - \mu'_{13}(q_1^2 + q_2^2)q_6^2 - \mu'_{31}(q_4^2 + q_5^2)q_3^2 \\
&- t'_{11}(P_1^2q_1^2 + P_2^2q_2^2) - t'_{33}P_3^2q_3^2 - t'_{12}(P_1^2q_2^2 + P_2^2q_1^2) - t'_{13}(P_1^2 + P_2^2)q_3^2 - t'_{31}(q_1^2 + q_2^2)P_3^2 \\
&- t_{44}P_1P_2q_1q_2 - t'_{44}(P_1P_3q_1q_3 + P_2P_3q_2q_3) \\
&- \kappa'_{11}(P_1^2q_4^2 + P_2^2q_5^2) - \kappa'_{33}P_3^2q_6^2 - \kappa'_{12}(P_1^2q_5^2 + P_2^2q_4^2) - \kappa'_{13}(P_1^2 + P_2^2)q_6^2 - \kappa'_{31}(q_4^2 + q_5^2)P_3^2 \\
&- \kappa_{44}P_1P_2q_4q_5 - \kappa'_{44}(P_1P_3q_4q_6 + P_2P_3q_5q_6) \\
&- \frac{\zeta_{44}\lambda_{44}}{C_{44}}(q_2q_3q_5q_6 + q_1q_3q_4q_6) + \frac{(C_{11} + 2C_{12})(C_{11} - C_{12})}{C_{11}}\varepsilon_S^2
\end{aligned} \qquad (A8)$$

where the ' sign represents the renormalized coefficients with thin film boundary condition, i.e.

$$\alpha_1'(T) = \alpha_1(T) - (g_{11} + g_{12} - \frac{2C_{12}}{C_{11}}g_{12})\varepsilon_S, \qquad \alpha_3'(T) = \alpha_1(T) - (2g_{12} - \frac{2C_{12}}{C_{11}}g_{11})\varepsilon_S,$$

$$\alpha_{11}' = \alpha_{11} + \alpha_{12} - \frac{g_{12}^2}{2C_{11}}, \qquad \alpha_{33}' = \alpha_{11} + \alpha_{12} - \frac{g_{11}^2}{2C_{11}},$$

$$\alpha_{12}' = 2\alpha_{11} - \frac{g_{12}^2}{C_{11}}, \qquad \alpha_{13}' = 2\alpha_{11} - (\frac{g_{11}g_{12}}{C_{11}} + \frac{g_{44}^2}{2C_{44}}),$$

$$\beta_1'(T) = \beta_1(T) - (\lambda_{11} + \lambda_{12} - \frac{2C_{12}}{C_{11}}\lambda_{12})\varepsilon_S, \qquad \beta_3'(T) = \beta_1(T) - (2\lambda_{12} - \frac{2C_{12}}{C_{11}}\lambda_{11})\varepsilon_S,$$

$$\beta_{11}' = \beta_{11} + \beta_{12} - \frac{\lambda_{12}^2}{2C_{11}}, \qquad \beta_{33}' = \beta_{11} + \beta_{12} - \frac{\lambda_{11}^2}{2C_{11}},$$

$$\beta_{12}' = 2\beta_{11} - \frac{\lambda_{12}^2}{C_{11}}, \qquad \beta_{13}' = 2\beta_{11} - (\frac{\lambda_{11}\lambda_{12}}{C_{11}} + \frac{\lambda_{44}^2}{2C_{44}}),$$

$$\gamma_1'(T) = \gamma_1(T) - (\varsigma_{11} + \varsigma_{12} - \frac{2C_{12}}{C_{11}}\varsigma_{12})\varepsilon_S, \qquad \gamma_3'(T) = \gamma_1(T) - (2\varsigma_{12} - \frac{2C_{12}}{C_{11}}\varsigma_{11})\varepsilon_S,$$

$$\gamma_{11}' = \gamma_{11} + \gamma_{12} - \frac{\varsigma_{12}^2}{2C_{11}}, \qquad \gamma_{33}' = \gamma_{11} + \gamma_{12} - \frac{\varsigma_{11}^2}{2C_{11}}, \tag{A9}$$

$$\gamma_{12}' = 2\gamma_{11} - \frac{\varsigma_{12}^2}{C_{11}}), \qquad \gamma_{13}' = 2\gamma_{11} - (\frac{\varsigma_{11}\varsigma_{12}}{C_{11}} + \frac{\varsigma_{44}^2}{2C_{44}}),$$

$$\mu_{11}' = \mu_{11} + \frac{\varsigma_{12}\lambda_{12}}{C_{11}}, \quad \mu_{33}' = \mu_{11} + \frac{\varsigma_{11}\lambda_{11}}{C_{11}}, \quad \mu_{12}' = \mu_{12} + \frac{\varsigma_{12}\lambda_{12}}{C_{11}},$$

$$\mu_{13}' = \mu_{12} + \frac{\varsigma_{11}\lambda_{12}}{C_{11}}, \quad \mu_{31}' = \mu_{12} + \frac{\varsigma_{12}\lambda_{11}}{C_{11}},$$

$$t_{11}' = t_{11} + \frac{g_{12}\lambda_{12}}{C_{11}}, \quad t_{33}' = t_{11} + \frac{g_{11}\lambda_{11}}{C_{11}}, \quad t_{12}' = t_{12} + \frac{\varsigma_{12}\lambda_{12}}{C_{11}},$$

$$t_{13}' = t_{12} + \frac{g_{12}\lambda_{11}}{C_{11}}, \quad t_{31}' = t_{12} + \frac{g_{11}\lambda_{12}}{C_{11}}, \quad t_{44}' = t_{44} + \frac{g_{44}\lambda_{44}}{C_{44}},$$

$$\kappa_{11}' = \kappa_{11} + \frac{g_{12}\varsigma_{12}}{C_{11}}, \quad \kappa_{33}' = \kappa_{11} + \frac{g_{11}\varsigma_{11}}{C_{11}}, \quad \kappa_{12}' = \kappa_{12} + \frac{\varsigma_{12}g_{12}}{C_{11}},$$

$$\kappa_{13}' = \kappa_{12} + \frac{\varsigma_{11}g_{12}}{C_{11}}, \quad \kappa_{31}' = \kappa_{12} + \frac{\varsigma_{12}g_{11}}{C_{11}}, \quad \kappa_{44}' = \kappa_{44} + \frac{\varsigma_{44}g_{44}}{C_{44}}.$$